\documentclass[mathleft
]{an}
\usepackage{graphicx}
\usepackage{times}
\begin{document}

\Pagespan{1}{}
\Yearpublication{2012}%
\Yearsubmission{2012}%
\Month{99}%
\Volume{999}%
\Issue{99}%

\title{Variability survey in the young open cluster IC 1805}

\author{D. Mo\'zdzierski\thanks{Corresponding author:
  \email{mozdzierski@astro.uni.wroc.pl}\newline}
}
\titlerunning{Variability survey in the young open cluster IC 1805}
\authorrunning{D.\,Mo\'zdzierski}
\institute{Astronomical Institute, University of Wroc{\l}aw, Kopernika 11, 51-622 Wroc{\l}aw, Poland
}

\received{October 2012}
\accepted{October 2012}
\publonline{November 2012}


\abstract{
  We present preliminary results of the photometric variability survey in the very young open cluster IC 1805.
  We found more than 70 variable stars in the field, including pulsating stars and a large sample of most likely
  pre-main sequence stars.}

\maketitle

\section{Introduction}
The variability survey in the young open cluster IC\,1805 is a part of the ongoing program of searching for early-type variable stars
in open clusters of the northern sky conducted in Wroc{\l}aw (see Jerzykiewicz et al.~2011 and references therein).
We are particularly interested in pulsating stars such as $\beta$ Cephei and SPB. These stars are very promising targets for asteroseismology.
In particular, open clusters rich in these variables can be used for so called ensemble asteroseismology (Saesen et al.~2010).

\begin{figure}
\centering
\includegraphics[width=80mm,height=69mm]{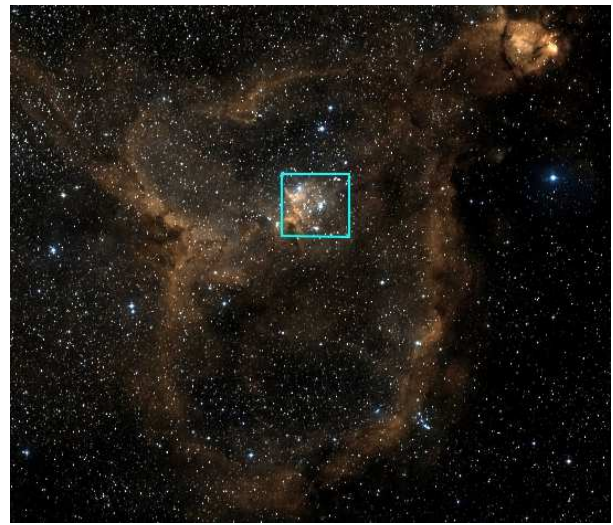}
\caption{DSS image of the Heart Nebula IC\,1805 and its central open cluster. The observed field is marked
by the cyan square.}
\label{field}
\end{figure}

The open cluster IC\,1805 is located in the Perseus spiral arm of our Galaxy in the centre of Cas OB6 association and the
molecular cloud W4. The surrounding nebula, named also IC\,1805, has a nickname Heart Nebula (Fig.~\ref{field}). The cluster is
very young; its age is estimated for only se\-ve\-ral Myr and distance, for about 2\,--\,2.4 kpc
(Ishida 1968; Joshi \& Sagar 1983; Guetter \& Vrba 1989; Massey, John\-son \& DeGioia-Eastwood 1995;
Sung \& Lee 1995). It is also heavily reddened and the reddening is not uniform across the cluster.
There was no thorough variability survey in the cluster up to date. The NSV catalog (Samus et al.~2010) lists only eight suspected variables 
falling into the field of view of our observations.


\section{Observations and reductions}
\begin{figure}
\centering
\includegraphics[width=80mm,height=77.6mm]{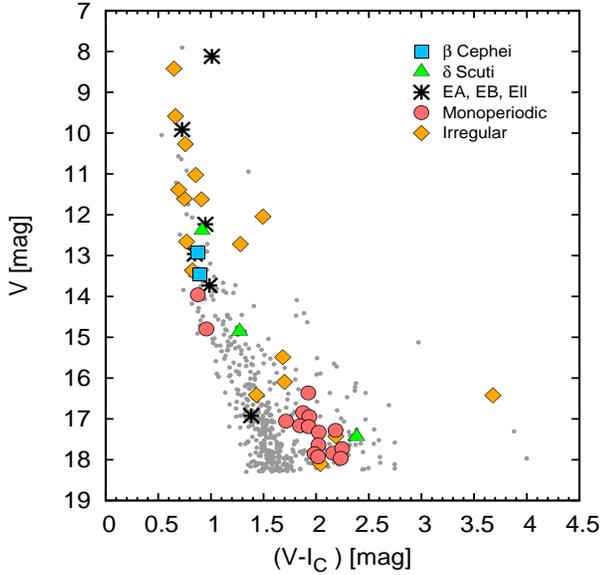}
\caption{$V$ vs. $(V-I_{\rm{C}})$ colour-magnitude diagram for stars in the observed field. All detected variables are
divided into five groups that are shown with different symbols and labeled. }
\label{cmdvar}
\end{figure}

All observations were carried out in Bia{\l}k\'ow Observatory
(University of Wroc{\l}aw) between 2007 and 2010. We used a 60-cm
Cassegrain telescope equipped with the Andor Tech.~DW 432-BV
CCD camera covering a 13$^{\prime}\times$\,12$^{\prime}$ field of view. We have
acquired almost ten thousands CCD frames during 44 observing nights. They
were taken through three filters, $B$, $V$ and $I_{\rm{C}}$. The exposure times ranged
from 10 to 100~s. The observed field of IC\,1805 is shown in Fig.~\ref{field}.
\newline
\indent
The observations were calibrated in a standard way 
which included bias and dark subtraction and 
correction for inhomogeneous sensitivity using flat-field frames. The aperture and profile
magnitudes calculated by means of the {\sc DAOPHOT II} package (Stetson 1987) were
used to derive differential magnitudes that were subsequently used in the variability search.
\newline
\indent
The instrumental $BVI_{\rm C}$ photometry
for all 1511 detected stars was transformed to the standard system using 
photometry published by Sung \& Lee (1995). The resulting colour-magnitude diagram is
shown in Fig.~\ref{cmdvar}. As can be seen, the main sequence of the cluster is
slightly smeared due to variable reddening.

\section{Analysis and results}
The search for variable stars was based on the analysis of differential magnitudes by means of a Fourier 
amplitude spectrum and eye inspection of light curves and calculation of phase diagrams.
The search has been done using the most numerous $I_{\rm C}$-filter observations. The
periodograms were calculated in the range between 0 and 80~d$^{-1}$.
In the first step, the identification of variable stars was made automatically. The light
curves and Fourier spectra of variable candidates were then inspected by eye.
In this way, some non-periodic variables were also identified. For multiperiodic stars,
we also carried out a prewhitening procedure in order to derive frequencies and amplitudes of
all significant signals.

\begin{figure}
\centering
\includegraphics[width=60.1mm,height=83mm,angle=270]{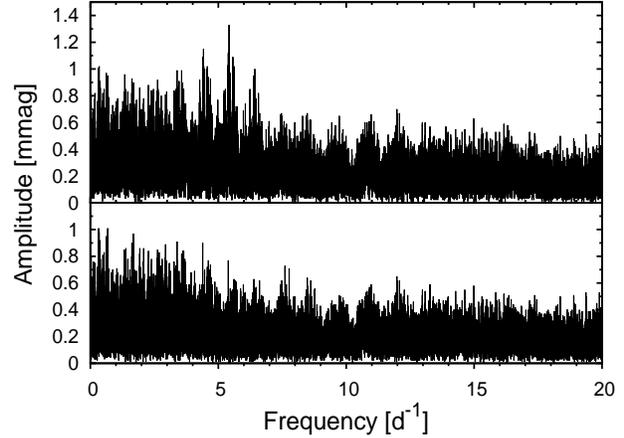}
\caption{Top: Fourier periodogram of one of the two $\beta$ Cephei variables we found. Bottom:
the same, after prewhitening with the dominant frequency. }
\label{betac}
\end{figure}

\begin{figure}
\centering
\includegraphics[width=58.1mm,height=83mm,angle=270]{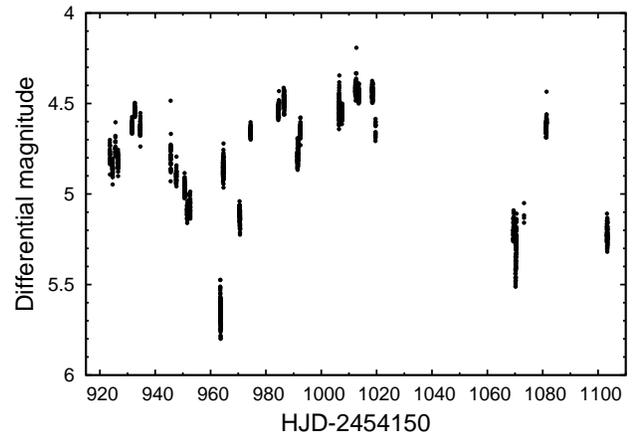}
\caption{The $I_{\rm C}$-filter light curve of one of UX Ori-type candidates in IC\,1805.}
\label{uxori}
\end{figure}

The search resulted in the detection of 71 variables. Only four of them were suspected
to be variable before our study, the remaining 67 are new discoveries. We found eight pulsating
stars in the observed field: two $\beta$ Cephei and six $\delta$ Scuti type stars. The two $\beta$ Cephei stars and one $\delta$ Scuti
star are likely members of IC\,1805. 
If this is the case, the $\delta$~Scuti star might be still in the pre-main sequence phase of evolution. The periodogram of one (low-amplitude) $\beta$ Cephei 
star is shown in Fig.\,\ref{betac}.
In addition, we found 8 stars showing variability caused by eclipses or ellipsoidal effects, 24 irregular and 31
monoperiodic variables. The monoperiodic variables occupy mostly the region where pre-main sequence members of IC\,1805 are expected (Fig.~\ref{cmdvar}).
Their variablity is probably related to rotation. The light curves of some of the irregular variables resemble those of UX Orionis stars
in which the variability is attributed to the changes in the circumstellar environment. An example light curve for one of IC\,1805 stars 
is shown in Fig.\,\ref{uxori}.

\acknowledgements

We are indebted to Prof.\,A.\,Pigulski and
Dr.\,Z.\,Ko{\l}aczkowski for their help and discussions.
We want to thank D.\,Drobek, M.\,Jurecki, G.\,Kopacki,
G.\,Michalska, P.\,Nowak, and M.\,St\c{e}\'slicki
for making some observations of the cluster.
The work was supported by the NCN grant 2011/03/B/ST9/02667.

\newpage

\end{document}